\documentclass[10pt,conference]{IEEEtran}
\usepackage{cite}
\usepackage[dvips]{color}
\usepackage{tabu}
\usepackage{comment}
\usepackage{todonotes}
\usepackage{epsf}
\usepackage{epsfig}
\usepackage{times}
\usepackage{epsfig}
\usepackage{graphicx}
\DeclareRobustCommand{\bbone}{\text{\usefont{U}{bbold}{m}{n}1}}
\usepackage{mathtools}
\usepackage{mathrsfs}
\usepackage{amssymb,amsmath}
\usepackage{amsmath}
\usepackage{amsfonts}
\usepackage{pdfpages}
\usepackage{epstopdf}
\usepackage{algorithmicx}  
\usepackage[algo2e]{algorithm2e} 
\usepackage{cleveref}
\usepackage{dsfont}
\usepackage{lettrine} 
\usepackage{epsfig,algorithm,algpseudocode,amsthm,url}
\usepackage[justification=raggedright]{caption}
\usepackage{subcaption}
\usepackage{bbm}
\usepackage[utf8]{inputenc}
\usepackage{csquotes}
\usepackage{verbatim}
\usepackage{setspace}	
\usepackage[T1]{fontenc}
\usepackage{textcomp}
\usepackage{tabularx}
\usepackage{xpatch}
\DeclareMathOperator*{\argmax}{argmax}

\usepackage{stackengine}
\usepackage[nocomma]{optidef}

\IEEEoverridecommandlockouts 
\begin{document}
\title{\huge Semantic-Aware Collaborative Deep Reinforcement Learning Over Wireless Cellular Networks
\thanks{\small F. Lotfi and O. Semiari are with Department of Electrical and Computer Engineering,
University of Colorado, Colorado Springs, CO, 80918 USA (e-mail: flotfi@uccs.edu; osemiari@uccs.edu).

W. Saad is with Wireless@VT, Department of Electrical and
Computer Engineering, Virginia Tech, Blacksburg, VA, 24061 USA (e-mail: walids@vt.edu).} \vspace*{-0.4cm}}
\author{
\authorblockN{Fatemeh Lotfi, Omid Semiari, and Walid Saad}\vspace*{-0.9cm}}
\maketitle\vspace{-0.7cm}
\begin{abstract}
Collaborative deep reinforcement learning (CDRL) algorithms in which multiple agents can coordinate over a wireless network is a promising approach to enable future intelligent and autonomous systems that rely on real-time decision making in complex dynamic environments. Nonetheless, in practical scenarios, CDRL face many challenges due to heterogeneity of agents and their learning tasks, different environments, time constraints of the learning, and resource limitations of wireless networks. To address these challenges, in this paper, a novel semantic-aware CDRL method is proposed to enable a group of heterogeneous untrained agents with semantically-linked DRL tasks to collaborate efficiently across a resource-constrained wireless cellular network. To this end, a new heterogeneous federated DRL (HFDRL) algorithm is proposed to select the best subset of semantically relevant DRL agents for collaboration. The proposed approach then jointly optimizes the training loss and wireless bandwidth allocation for the cooperating selected agents in order to train each agent within the time limitation of its real-time task. 
Simulation results show the superior performance of the proposed algorithm compared to state-of-the-art baselines.  
\vspace{-0.1cm}
\end{abstract}
\section{Introduction} \vspace{-0.1cm}
Emerging wireless autonomous systems, such as smart factories and intelligent transportation systems, will require autonomous Internet-of-Things (IoT) agents (e.g., robots, vehicles, drones) with computing and learning capabilities to perform real-time decision-making in dynamic environments~\cite{6Gwalid}. 
Deep reinforcement learning (DRL) is viewed as a promising approach for training intelligent IoT agents to tackle complex tasks such as autonomous navigation.   
However, DRL methods can lead to slow convergence~\cite{Nagib2021slowconv}, overfitting problems~\cite{zhang2018study}, or sub-optimal performance due to poor exploration in complex environments~\cite{yuan2021multimodal}. These challenges limit the applications of DRL for real-time autonomous IoT services where convergence time, generalizability of the learning, and performance are all important factors. 

To address these limitations, recently, a body of work in~\cite{boloka2021knowledge,ren2018collaborative,kong2017collaborative,abdolshah2021new,tao2021repaint,gupta2017learning,lin2017collaborative,lai2020dual,teh2017distral,zhao2020robust} focused on \emph{collaborative DRL (CDRL)} solutions that enable agents to share their experiences (e.g., state trajectories and rewards) and collaboratively learn the optimal policy for their task. Nonetheless, in most practical scenarios, CDRL will require frequent data communications among agents over a wireless network, and, thus, the limitations of the wireless link (e.g., limited bandwidth) will directly impact the CDRL performance and accuracy.  
Moreover, enabling an effective CDRL among heterogeneous agents (as expected in IoT applications) is very challenging, due to dissimilarities of agents (e.g., different action spaces), environments, and diversity of DRL tasks. In the CDRL context, the heterogeneity of environments, modeled as Markov decision processes (MDPs), as well as agents and their tasks can be expressed in two main forms: 1) distinct DRL tasks that are conceptually similar (i.e., \emph{semantically related} tasks) or completely dissimilar~\cite{abdolshah2021new,tao2021repaint} 
and 2) distinct environments represented by different MDPs~\cite{lin2017collaborative,gupta2017learning}. 
Most existing works, such as in \cite{boloka2021knowledge,ren2018collaborative,kong2017collaborative}, study CDRL among homogeneous agents, i.e., agents with the same action space. Meanwhile, the works in \cite{lin2017collaborative,gupta2017learning,abdolshah2021new,tao2021repaint} consider more realistic scenarios by adopting knowledge transfer for CDRL among heterogeneous agents. 
While interesting, most of the prior art in~\cite{lin2017collaborative,gupta2017learning,abdolshah2021new,tao2021repaint} assumes that an expert agent is always available and DRL tasks are semantically similar -- strong assumptions that are often inapplicable to practical IoT scenarios.

To address the challenges of CDRL among untrained agents, recent works in \cite{lai2020dual,teh2017distral,zhao2020robust} investigate knowledge transfer through Kullback-Leibler divergence (KLD) regularization and policy distillation between homogeneous agents with similar DRL tasks \cite{lai2020dual}, with semantically different tasks \cite{teh2017distral}, or in different environments \cite{zhao2020robust}. Although the works in \cite{lai2020dual,teh2017distral,zhao2020robust} address knowledge transfer between untrained agents, the proposed methods are designed only for homogeneous agents. 
Furthermore, most of these works do not account for the underlying \emph{semantic relatedness} of DRL tasks across agents and only rely on traditional similarity metrics (e.g., action-state spaces) in MDP-based environments. In addition, most of existing federated RL (FRL) approaches (e.g., see~\cite{qi2021federated} and references therein) involve all agents in CDRL without accounting for the similarities of DRL tasks, or for the resource constraints of the wireless network. Hence, there is a need for novel CDRL techniques that can address these problems.

The main contribution of this paper is a novel CDRL approach that enables efficient knowledge transfer and collaboration among a set of heterogeneous untrained agents with semantically related DRL tasks over a resource-constraint wireless cellular network. To this end, we introduce a novel heterogeneous federated DRL (HFDRL) that first extracts the similarities of DRL tasks to choose an optimal subset of semantically-related DRL agents for collaboration. 
Then, based on the proposed agent-selection scheme, the developed algorithm jointly optimizes the collaborating agents' training loss and wireless bandwidth allocation to train each agent within the time constraint of its real-time task. 
Simulation results show that the proposed CDRL method can yield up to $83\%$ performance gains compared to the state-of-the-art algorithms that do not account for the semantic-relatedness of tasks or wireless constraints. \emph{To the best of our knowledge, this is the first work that proposes a semantic-aware CDRL among heterogeneous untrained agents over a wireless cellular network}.

The rest of this paper is organized as follows. Section \ref{sysmodl} presents the system model and the problem formulation for the CDRL over a wireless network. Section \ref{sem_DRL} presents the proposed HFDRL algorithm. Simulation results are presented in Sec. \ref{simulation} and conclusions are drawn in Sec. \ref{conclusion}. \vspace{-0.1cm}
\section{System Model}\label{sysmodl}
\begin{figure}[t!]
  \centering
    \includegraphics[width=8.3cm]{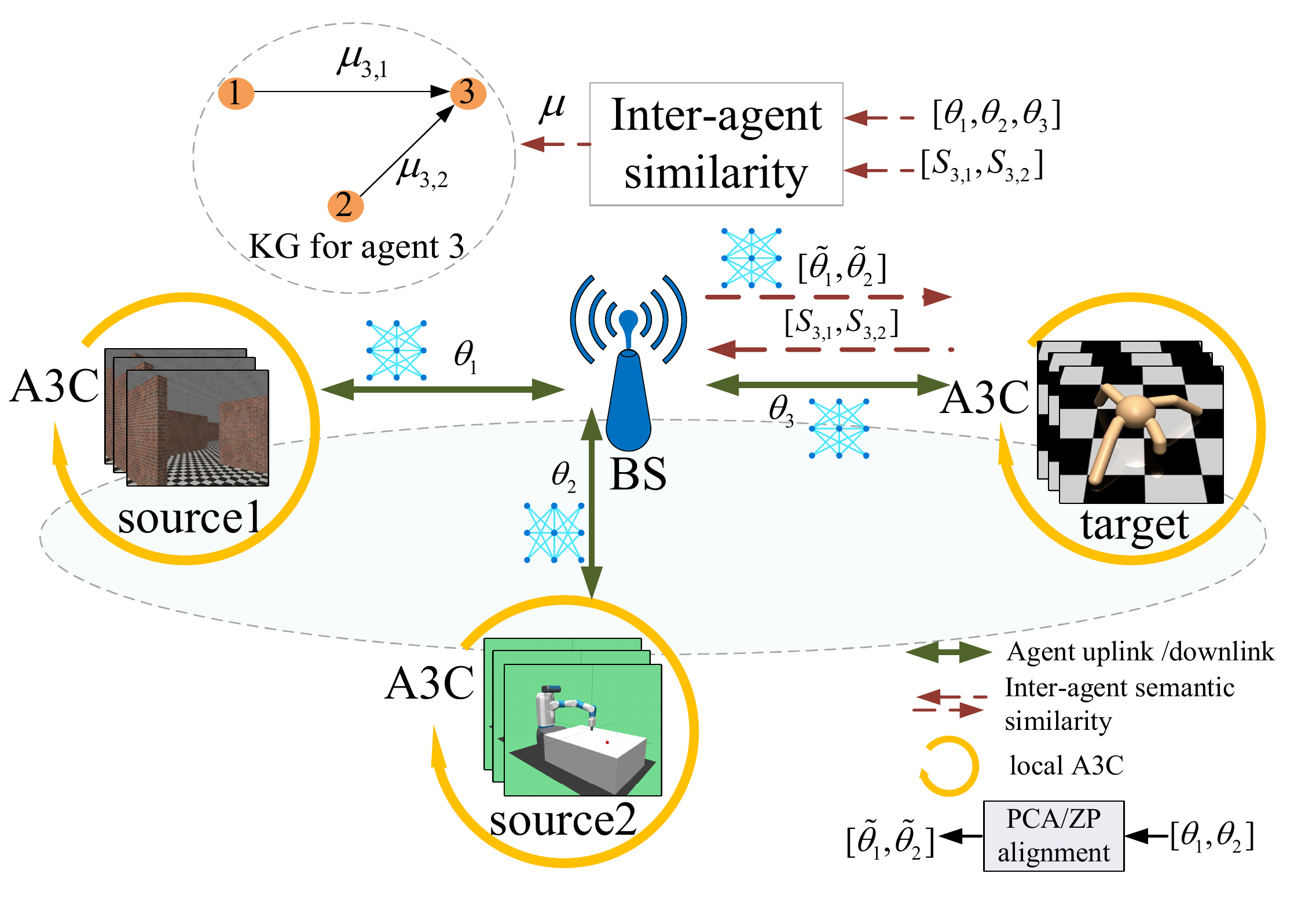}\vspace{-0.3cm}
    \caption{\small The proposed semantic-aware HFDRL scheme over a wireless cellular network. }\vspace{-0.2cm}
    \label{sys_graph}
\end{figure}

\vspace{-0.0cm}
Consider a wireless network having a set $\mathcal{N}$ of $N$ CDRL agents, as shown in Fig. \ref{sys_graph}. 
Each agent $n\in\mathcal{N}$ interacts with a distinct environment, formally characterized by an MDP and aims to perform its DRL task while collaborating with other agents over wireless links. 
The MDP for an agent $n$ is defined as a tuple $\langle \mathcal{S}_n,\mathcal{A}_n,T,r,\gamma \rangle$, where $\mathcal{S}_n$ and $\mathcal{A}_n$ are, respectively, the state space and the action space. In addition, $T(s_{n,t+1}|s_{n,t},a_{n,t})$ denotes the transition probability from the current state at time $t$, $s_{n,t} \in \mathcal{S}_n$, to the next state $s_{n,t+1}$, if an action $a_{n,t} \in \mathcal{A}_n$ is selected by the agent. Further, $r(s_{n,t},a_{n,t})$ is the agent's received reward after this transition. Also,  $0<\gamma< 1$ is a discount factor used to define the discounted return $R_{n,t} = \sum_{i=0}^{\infty}\gamma^i r_{n,t+i}$. 
The mapping from the state space to the action space at any given time $t$ is via the agent’s policy distribution $\pi_n(a_{n,t}|s_{n,t};\boldsymbol{\theta}_n)$ which is determined by a deep neural network (DNN) parametrized by $\boldsymbol{\theta}_n$. Given a policy $\pi_n$, the state-value and action-value functions are defined as $V_{\pi_n}(s_{n,t}) = \mathbb{E}_{\pi_n}[R_{n,t}\mid s_{n,t}]$ and $Q_{\pi_n}(s_{n,t},a_{n,t}) = \mathbb{E}_{\pi_n}[R_{n,t}\mid s_{n,t},a_{n,t}]$, respectively. 

\subsection{Collaborative Deep Reinforcement Learning Model}
Prior works in~\cite{lin2017collaborative,boloka2021knowledge,ren2018collaborative,kong2017collaborative,gupta2017learning,abdolshah2021new,tao2021repaint,zhao2020robust,lai2020dual,teh2017distral} have shown the effectiveness of CDRL, particularly for scenarios in which there is some degree of similarity in agents' experiences. As an example, consider a group of different autonomous robots that work in distinct environments, such as cleaning bots, delivery bots, hospitality bots, and autonomous drones. Although these agents have unique abilities and missions, the operation of the collective set of agents can be based on similar DRL tasks for scanning an area, clustering data, or finding the shortest path.

With this in mind, an agent $n \in \mathcal{N}$ can collaborate with other agents over a wireless network by sharing its local experience $\mathcal{M}_{n,t}$. In most prior works (e.g., see ~\cite{tao2021repaint} and ~\cite{gupta2017learning}), the local experience is defined as a tuple $\langle s_{n,t\:'}, a_{n,t\:'},\pi_n,r_{n,t\:'} \rangle$ for $0\leq t\:' \leq t$ that directly captures the agent's space exploration and received returns. However, this method is ineffective for heterogeneous agents and also can lead to substantial data traffic in large-scale scenarios. Therefore, this approach is not suitable for CDRL over wireless networks. In the proposed model, as shown in Fig. 1 and elaborated in Sec. III, $\mathcal{M}_{n,t}$ depends on the parameters of the policy network $\boldsymbol{\theta}_n$ as well as the \emph{semantic relationships} among DRL tasks. In this model, an arbitrary agent $n \in \mathcal{N}$ that requires training is called a \emph{target agent} while a subset of the remaining $N-1$ agents that help the target agent with its training are called \emph{source agents}. The source agents can be either experts (with pre-trained policies) or new learners (with some initial random policy). The DRL task of the target agent $n$ is to find an optimal policy $\pi_n^*$ (i.e., to train the DNN policy network and determine its $\boldsymbol{\theta}_n$) that maximizes its average discounted return, while leveraging the transferred knowledge by source agents. Next, we present the wireless model and, accordingly, formulate the CDRL problem cognizant of wireless resource constraints.\vspace{-0.1cm}
\subsection{Wireless Communication Model for CDRL }\vspace{-0.0cm}
To manage the concurrent transmissions of the agents' local experiences $\mathcal{M}_{n,t}$, $\forall n \in \mathcal{N}$, we consider an orthogonal frequency-division multiple access (OFDMA) scheme. Here, the base station (BS) serves as a coordinator that manages the data traffic and facilitates collaboration among CDRL agents. The data rate of an agent $n \in \mathcal{N}$ over an uplink frame $m$ will be\vspace{-0.4cm}
\begin{align}\label{urate}
    &c_{n,m}^\textrm{u} = B \sum_{k=1}^{K_\textrm{u}}e_{nk,m}^\textrm{u} \log\Big(1+\frac{p_n d_{n}^{-\eta} h_{nk,m}^\textrm{u}}{ \sigma^2}\Big),
\end{align}
where $B$ represents the resource block (RB) bandwidth and $K_\textrm{u}$ is the total numbers of RBs for uplink communications. In addition, $e_{nk,m}^\textrm{u} \in \{0,1\}$ is the uplink resource allocation variable with $e_{nk,m}^\textrm{u}=1$ indicating that RB $k$ is assigned to an agent $n$ at frame $m$, otherwise $e_{nk,m}^\textrm{u}=0$. 
$p_{n}$ is the transmit power of agent $n$ per RB and $d_{n}$ is the distance of the agent $n$ from the BS. Moreover, $\eta$ represents the path loss exponent, and $h_{nk,m}^\textrm{u}$ is the Rayleigh fading channel gain of RB $k$ for the link between agent $n$ and the BS. In \eqref{urate}, $\sigma^2$ represents the variance of the additive white Gaussian noise (AWGN). 
Similarly, the data rate for an agent $n\in \mathcal{N}$ at a downlink frame $m$ is given by \vspace{-0.2cm}
\begin{align}\label{drate}
    &c_{n,m}^{\textrm{d}} = B \sum_{k=1}^{K_\textrm{d}}e_{nk,m}^\textrm{d} \log\Big(1+\frac{p_{b} d_{n}^{-\eta} h_{nk,m}^\textrm{d}}{ \sigma^2}\Big),
\end{align}
where $e_{nk,m}^\textrm{d} \in \{0,1\}$ is the downlink resource allocation variable with $e_{nk,m}^\textrm{d} =1$ if RB $k$ is assigned to agent $n$ at frame $m$, otherwise $e_{nk,m}^\textrm{d}=0$. 
Also, $p_{b}$ is the BS transmit power per RB.
From \eqref{urate} and \eqref{drate}, and by considering $\beta$ as the size of data packets that encode the agent's local experience, the over-the-air transmission delay can be obtained from\vspace{-0.1cm}
\begin{align}\label{tdelay}
    &\tau_{n,m}^\textrm{i} = \frac{\beta}{c_{n,m}^\textrm{i}},
\end{align}
where $\textrm{i}\in\{\textrm{u},\textrm{d}\}$ indicates that \eqref{tdelay} is applied to both uplink and downlink.
Based on the transmission delay in \eqref{tdelay}, an agent can utilize the shared real-time experience by other agents, only if the information is received within a time period $\tau_{\text{th}}$.\vspace{-0.1cm}
\subsection{Problem Formulation}
Given the proposed CDRL model over the wireless network, we formulate a new optimization problem whose goal is to find an optimal policy for a target agent $n$ while accounting for the bandwidth constraints at any communications frame $m$, i.e., \vspace{-0.1cm}
\begin{subequations} 
\begin{align}\label{opt1}
 \argmax_{\boldsymbol{\theta}_{n},e_{nk,m}^\textrm{i}} & \hspace{0.7cm}  \mathbb{E}_{\pi_n^*}[R_{n,t}(\boldsymbol{\theta}_{n},e_{nk,m}^\textrm{i})\mid \mathcal{M}_t],\\
\text{s.t.,} 
& \hspace{0.7cm} \tau_{n,m}^\textrm{i} <\tau_{\text{th}} , \label{opt1_delay}\\
& \hspace{0.7cm} \sum_{n=1}^{N}\sum_{k=1}^{K_\textrm{i}} e_{nk,m}^\textrm{i} \leq K_\textrm{i}, \label{opt1_RB}\\
& \hspace{0.7cm} \sum_{n=1}^{N}e_{nk,m}^\textrm{i} \leq 1,\,\, \forall k\in K_\textrm{i}, \label{opt1_RB2}
\end{align}
\end{subequations}
where in \eqref{opt1_delay}-\eqref{opt1_RB2}, $ \textrm{i} \in \{\textrm{u},\textrm{d}\}$ indicates that the constraints 
are applied to both uplink and downlink, and $\boldsymbol{\theta}_n$ is DNN parameters of the target agent optimal policy $\pi_{n}^*$. Moreover, $\mathcal{M}_t=\bigcup\limits_{n\in \mathcal{N}}\mathcal{M}_{n,t}$. \eqref{opt1_delay} constrains the over-the-air transmission delay at uplink and downlink to be less than the time constraint $\tau_{\text{th}}$. Here, without loss of generality, we let $\tau_{\text{th}}$ be the time slot duration to ensure timely collaboration between real-time agents over both uplink and downlink. The constraint in \eqref{opt1_RB} guarantees that the total allocated RBs must be less than or equal to the total available RBs. Moreover, \eqref{opt1_RB2} ensures orthogonal allocation of RBs to agents.  
Meeting these feasibility constraints can be challenging for real-time CDRL in large networks, due to the limited available bandwidth. 
Furthermore, evaluating whether the experience of other agents would be useful in optimizing the policy for a target agent is challenging due to the different environments and tasks across different agents. Hence, policy training for the CDRL agents must be optimized jointly with the source agent selection and wireless resource allocation. 
Given these challenges, the problem in \eqref{opt1}-\eqref{opt1_RB2} cannot be solved via existing CDRL methods such as knowledge distillation \cite{lin2017collaborative,lai2020dual,zhao2020robust,teh2017distral} to transfer the source agents' knowledge to the target agent. 
Moreover, existing CDRL methods cannot be directly applied here as they do not account for the wireless and real-time constraints of knowledge sharing among agents. \vspace{-0.0cm}
\section{Semantic-Aware Heterogeneous Federated Deep Reinforcement Learning}\label{sem_DRL}\vspace{-0.1cm}
Next, we develop a novel CDRL approach to address these challenges and solve the problem in \eqref{opt1}-\eqref{opt1_RB2}.\vspace{-0.1cm}

\subsection{Heterogeneous Federated DRL}\vspace{-0.0cm}
To enable transfer learning among heterogeneous agents and over a resource-limited wireless network, we develop a new method, called HFDRL, that builds on heterogeneous federated learning (HeteroFL) along with a new agent-selection scheme that is cognizant of DRL task similarities, so as to solve the problem in \eqref{opt1}-\eqref{opt1_RB2}. 
In the proposed HFDRL approach, the local model parameters of each agent's policy network are considered as a subset of the global model parameters, i.e., $\boldsymbol{\theta}_n\subseteq\boldsymbol{\theta}_g$, where $\boldsymbol{\theta}_g$ in each hidden layer is represented as a matrix with dimensions $x_g$ and $y_g$. Using this method, local models are categorized into $L$ different levels of computational complexity compared to the global model. That is, $x_n^l = \zeta^{l-1}x_g$ and $y_n^l = \zeta^{l-1}y_g$ are the dimensions of the level $l$ local models, and $\zeta$ captures the local models' shrinkage ratio. 
HeteroFL performs model aggregation as~\cite{diao2020heterofl} \vspace{-0.0cm}
\begin{align}\label{g_hetfed}
    \boldsymbol{\theta}_g = \boldsymbol{\theta}_n^L \cup (\boldsymbol{\theta}_n^{L-1}\setminus \boldsymbol{\theta}_n^L) \cup \cdots \cup (\boldsymbol{\theta}_n^1 \setminus \boldsymbol{\theta}_n^2), \,\, \forall n \in \mathcal{N},
\end{align}
where, $\boldsymbol{\theta}_n^{L-1}\setminus \boldsymbol{\theta}_n^L$ represents the set of elements that are present in  $\boldsymbol{\theta}_n^{L-1}$ but not in $\boldsymbol{\theta}_n^L$. Accordingly, the parameters of the policy network for an agent $n$ can be obtained from $\boldsymbol{\theta}_n^l = \boldsymbol{\theta}_g[1:x_n^l,1:y_n^l]$, which is a sub-matrix with dimensions $x_n^l$ and $y_n^l$. 
Further, as mentioned in Sec. I, the experience (or equivalently, the policy network) of a random source agent may not be relevant to a target agent due to heterogeneous tasks and environments. To this end, in contrast with existing HeteroFL methods (e.g., in \cite{diao2020heterofl}), the goal in our HFDRL is to minimize a global loss function that is defined as a \emph{weighted} average of the local loss functions from collaborating agents where weights are defined based on the similarities of tasks. That is, to solve the problem in \eqref{opt1}-\eqref{opt1_RB2} for a target agent $n$ in a general CDRL scenario with heterogeneous agents and tasks, we reformulate the problem as \vspace{-0.2cm}
\begin{subequations} 
\begin{align}\label{opt2}
\min_{\boldsymbol{\theta}_{g},e_{nk,m}^i} & \hspace{0.7cm}  F(\boldsymbol{\theta}_{g})=\sum_{n'=1}^{N}W_{n,n'}f_{n'}(\boldsymbol{\theta}_{n'}),\\
\text{s.t.,} 
& \hspace{0.7cm} \eqref{opt1_delay}-\eqref{opt1_RB2}. \label{opt2_2}\vspace{-0.3cm}
\end{align}
\end{subequations}
where $f_{n'}(\boldsymbol{\theta}_{n'})$ represents the loss function of the local model for agent $n'\in \mathcal{N}$. Considering policy gradient (PG) DRL algorithms \cite{lin2017collaborative}, the loss function is defined as\vspace{-0.1cm}
\begin{align}\label{lossf}
 f_n(\boldsymbol{\theta}_{n})=-A_{\pi_n}(s_{n},a_{n}) \log (\pi_n(a_{n}|s_{n};\boldsymbol{\theta}_n)),
\end{align}
where $A_{\pi_n}(s_{n},a_{n}) =Q_{\pi_n}(s_{n},a_{n})-V_{\pi_n}(s_{n})$ represents the action advantage function. Here, by considering asynchronous advantage actor critic (A3C) model for each agent, maximizing the advantage value would be equivalent to the highest return value in \eqref{opt1}. Therefore, solving \eqref{opt2}-\eqref{opt2_2} will yield the optimal solution for the original problem in \eqref{opt1}-\eqref{opt1_RB2}. 
In \eqref{opt2}, $W_{n,n'}$ controls the influence of heterogeneous source agents $n'$ in policy training for a target agent $n$. We can observe that the convergence of \eqref{opt2} is affected by the weight of each agent. To determine the optimal weights, we develop a new similarity metric between an arbitrary target agent and source agents. Using this similarity metric, a subset of source agents will be selected to participate in the HFDRL so as to 1) prevent irrelevant agent experiences to impact the policy for the target agent and 2) minimize the wireless data traffic and meet the delay constraints for the HFDRL. \vspace{-0.2cm}
\subsection{Proposed Similarity Metric and Knowledge Graph}
Finding a proper similarity metric and deriving the weights, $W_{n,n'}$, are critical to ensure a \emph{positive} collaboration among agents. In such a case, after collaboration, the target agent must achieve a higher average reward compared to a case in which learning is done from scratch without any collaboration with other agents. Therefore, measuring inter-agent similarity is important to avoid a negative collaboration between agents. 
To this end, we define an appropriate similarity metric that exploits inter-agent relatedness, both structurally and semantically. 
While a limited number of recent works consider structural similarity \cite{visus2021taxonomy}, or semantic similarity \cite{tao2021repaint} in isolation, as we elaborate next, none of these metrics alone are sufficient to effectively capture similarity between heterogeneous tasks and environments. 
Before introducing the proposed similarity metric, we define the structural and semantic similarities in the context of CDRL. 

\emph{Inter-agent structural similarity:} 
Given that the performance of DRL agents and their operation in an environment is determined by their trained policies, structural similarities between their policy networks can capture the similarity of agents \cite{visus2021taxonomy}. Here, we use cosine similarity as a heuristic metric to define the structural similarity $C_{n,n'}$ between a target agent $n$ and a source agent $n'$ \cite{visus2021taxonomy}:\vspace{-0.2cm}
\begin{align}\label{cosine}
    C_{n,n'}=1-\frac{\langle \boldsymbol{\theta}_n,\boldsymbol{\theta}_{n'} \rangle}{\|\boldsymbol{\theta}_n\| \|\boldsymbol{\theta}_{n'}\|},
\end{align}
where, $\langle \boldsymbol{\theta}_n,\boldsymbol{\theta}_{n'} \rangle$ represents inner product of $\boldsymbol{\theta}_n$ and $\boldsymbol{\theta}_{n'}$ as the parameters of the policy networks for the target agent $n$ and source agent $n'$, respectively. 
As stated before, the structural similarity metric alone may not be enough to optimize the CDRL. In fact, even if this parameter shows a low similarity between agents, one cannot conclude with certainty that agents' participation in the HFDRL may lead to a negative collaboration. That is because heterogeneous agents may train structurally dissimilar policy networks while their underlying DRL tasks are semantically related. \vspace{-0.0cm}

\emph{Inter-agent semantic relatedness:}
The effectiveness of CDRL is highly dependent on the similarity of underlying learning tasks pursued by the agents. In this regard, semantic relatedness in the transfer knowledge domain can quantify how much the transferred knowledge from a source agent might help the target agent in finding its optimal policy. 
A source agent $n'$ (trained or untrained) with a policy $\pi_{n'}$ is said to have the highest \emph{semantic relatedness} with the target agent, if it can extract the maximum average return value from a target environment in a limited number of training episodes. Hence, inspired by the work in \cite{tao2021repaint}, we define the semantic relatedness $S_{n,n'}$ between a source agent $n'$ and the target agent $n$ by the return value of the target agent under the source agent's policy: \vspace{-0.5cm}
\begin{align}\label{semantic}
    S_{n,n'} =\frac{1}{N_e}\sum_{i=1}^{N_e} R_{n,t_i}(\pi_{n'}), 
\end{align}
where $N_e$ is the number of training steps.    
Although the inter-agent semantic relatedness in \eqref{semantic} provides useful information about the relationship between tasks, this metric alone is not enough as it builds up return values over limited training steps. Hence, results may not be accurate, particularly for a complex target environment with a large state space. 

Therefore, for the proposed HFDRL scheme, we define a new similarity metric between a source agent $n'$ and the target agent $n$ as \vspace{-0.4cm}
\begin{align}\label{mu}
    \mu_{n,n'} = \alpha C_{n,n'} + (1-\alpha) S_{n,n'},
\end{align}
where $\alpha$ is a scalar variable that controls the importance of each of the two elements in the similarity measure. 
Due to the heterogeneity of agents, to compute $\mu_{n,n'}$ in \eqref{mu}, the source agent's policy network parameters $\boldsymbol{\theta}_{n'}$ must be aligned to the dimensions of the target agent's DNN. Therefore, we define $G(\boldsymbol{\theta}_n,\boldsymbol{\theta}_{n'})$ as an alignment function that utilizes a principal component analysis (PCA) method \cite{goodfellow2016deep} for compression or dimensional reduction, and a zero-padding (ZP) method for expansion of the DNN parameters: \vspace{-0.1cm}
\begin{align}
    G(\boldsymbol{\theta}_n,\boldsymbol{\theta}_{n'}) =  \begin{cases}
     \tilde{\boldsymbol{\theta}}_{n'} = \text{PCA}(\boldsymbol{\theta}_{n'}), & \text{for compression },\\
      \Tilde{\boldsymbol{\theta}}_{n'} = \text{ZP}(\boldsymbol{\theta}_{n'}), & \text{for expansion}.    
    \end{cases}
\end{align}
PCA is a common method for reducing dimensions. It is a projection-based approach for transforming data by projecting it into a lower dimensional encoded format \cite{goodfellow2016deep}. Using the developed similarity metric $\mu_{n,n'}$ in \eqref{mu}, we define a knowledge graph (KG) to score and represent the relationship between the agents based on the extracted similarity information. The KG is defined as a weighted directed graph with $N$ agents in $\mathcal{N}$ as its vertices and $\mu_{n,n'}$ as the weight of the edge from $n$ to $n'$. 
As DRL agents' get trained over time, KG will present more accurate information about inter-agent similarities. 

Using the KG, the BS can efficiently select a subset of source agents that can contribute to the training of each target agent in the HFDRL. That is, using the KG, the BS sets the model averaging weights $W_{n,n'}$ in \eqref{opt2} for training a target agent $n$ to: \vspace{-0.3cm}
\begin{align}\label{weight}
     W_{n,n'} = \bbone_{\mu_{n,n'} \geq \lambda}\Bigg[\frac{\mu_{n,n'}}{\sum_{n'\in \mathcal{N}_n}\mu_{n,n'}}\Bigg] , \,\, \forall n' \in \mathcal{N}_n,
\end{align}
where $\mathcal{N}_n$ is the set of all agents in $\mathcal{N}$ except the target agent $n$. Moreover, $\bbone_{\mu_{n,n'} \geq \lambda}=1$, if $\mu_{n,n'} \geq \lambda$, otherwise  $\bbone_{\mu_{n,n'} \geq \lambda}=0$. \eqref{weight} ensures that if the similarity measure between a source agent $n'$ and the target agent $n$ is small, $n'$ will not contribute to the policy learning by $n$, and thus, the proposed agent-selection method can prevent negative collaborations.\vspace{-0.1cm}
\subsection{Proposed HFDRL Algorithm}
Algorithm 1 summarizes the proposed semantic-aware HFDRL algorithm to solve the problem in \eqref{opt2}-\eqref{opt2_2}. 
The algorithm proceeds as follows to output the trained policy for each target agent. 
In Step 1, each agent $n \in \mathcal{N}$ trains/updates its local policy network for one training episode, then sends its model parameter $\boldsymbol{\theta}_{n}$ to the BS. In Step 2, the BS aligns the dimensions of the models from source agents according to the target model dimension, then calculates $C_{n,n'}, \forall n'\in\mathcal{N}_n$ based on \eqref{cosine}, and sends the $\tilde{\boldsymbol{\theta}}_{n'}$ to the target agent. Then in Step 3, the target agent utilizes the source agents aligned parameters $\tilde{\boldsymbol{\theta}}_{n'}$, calculates the semantic similarity $S_{n,n'}$ by averaging the obtained return values after $N_e$ episodes, according to \eqref{semantic}, and sends the results back to the BS. In Step 4, the BS calculates $\mu_{n,n'}, \forall n'\in\mathcal{N}_n$  based on \eqref{mu}, generates the KG, measures the appropriate weight for each agents based on \eqref{weight}, and selects the related agents to the target agent. In Step 5, the BS solves the bandwidth allocation problem and determines $e_{nk,m}^\textrm{u}$ and $e_{nk,m}^\textrm{d}$ via standard techniques, such as convex approximation methods. Finally, in Step 6, the BS calculates the global aggregation model parameters $\boldsymbol{\theta}_g$ based on the HeteroFL in \eqref{g_hetfed}, then computes the target agent's local model parameters $\boldsymbol{\theta}_n^L$ and sends them to the target agent. The algorithm terminates once the loss function in \eqref{lossf} does not change much over two consecutive training steps (i.e., convergence) or after $M_{max}$ communication rounds. \vspace{-0.1cm}
\begin{algorithm}[t!]
\SetAlgoLined
\textbf{Input}: $\boldsymbol{\theta}_{n}\,$ $\forall n\in\mathcal{N}$, $N$, $\alpha$, target agent $n$, $K_\textrm{u}$, $K_\textrm{d}$, $\tau_{th}$, $\lambda$, $M_{max}$.   \\
\While{the termination condition is met}{
\text{1:} Each agent $n \in \mathcal{N}$ sends $\boldsymbol{\theta}_{n}$ to the BS.

\text{2:} The BS aligns the dimensions of $\boldsymbol{\theta}_{n'}$ and calculate $C_{n,n'}$ based on \eqref{cosine}.

\text{3:} The target agent utilizes $\tilde{\boldsymbol{\theta}}_{n'}$, calculates $S_{n,n'}$ after $N_e$ episodes according to \eqref{semantic}, and send it to the BS.

\text{4:} The BS calculates $\mu_{n,n'}$  based on \eqref{mu}, and select the related agents based on the KG and \eqref{weight}.

\text{5:} the BS optimally allocates RBs and determines $e_{nk,m}^\textrm{u}$ and $e_{nk,m}^\textrm{d}$. 

\text{6:} The BS calculates the global aggregation model parameters based on HeteroFL as \eqref{g_hetfed}, computes $\boldsymbol{\theta}_{n}^L$ and sends it to the target agent $n$.
}
\textbf{Output}: $\boldsymbol{\theta}_{n}^L,e_{nk,m}^\textrm{u},e_{nk,m}^\textrm{d}\,\,\,$ for the target agent $n\in\mathcal{N}$. \\
\caption{The proposed HFDRL algorithm}\vspace{-0.1cm}
\label{alg1}
\end{algorithm}\vspace{-0.0cm}
\section{Simulation Results}\label{simulation}
To evaluate the performance of the proposed algorithm, we consider $N=10$ DRL agents in two different classic control environments: Cartpole-v0 and Acrobot-v1 of Open-AI Gyms \cite{openai2016}. The Cartpole-v0 consists of a cart that can be moved to the left or right and a pole positioned vertically above it. The goal is to keep the pole straight. The state-space of the Cartpole-v0 is a 1-dimensional array consisting of four floating values, representing the horizontal position of the cart, velocity, angle of the pole, and angular velocity of the cart. The action space is discrete, whereby potential actions are 0 and 1 to push the cart, respectively, to the left and right. 
The second environment is Acrobot-v1 which is a two-link pendulum that only activates at the second joint. The two links initially point downward, and the aim is to swing the end-effector at least one link over the base. The state space consists of the $\sin(.)$ and $\cos(.)$ of the two rotational joint angles along with the joint angular velocities. The action space is to deliver +1, 0, or -1 torque to the joint between the two pendulum links. 
\begin{table}[t!] 
	\footnotesize
	\centering
	\caption{\vspace*{-0cm} System parameters }\vspace{-0.2cm}
	\begin{tabular}{|>{\centering\arraybackslash}m{1.2cm}|>{\centering\arraybackslash}m{2cm}|>{\centering\arraybackslash}m{1.2cm}|>{\centering\arraybackslash}m{1.4cm}|}
		\hline
		\bf{Parameter} &\bf{Value } &\bf{Parameter }&\bf{Value } \\
		\hline
		Subcarrier spacing & 60 kHz & Total bandwidth & 100 MHz  \\
		\hline
		$B$  & 720 kHz & $p_b$ & 56 dBm \\
		\hline
		$K_\textrm{u}$,$K_\textrm{d}$ & 135 &$\lambda$ & $[0.1,0.5,1.0]$ \\
		\hline
		 $p_n$ & 12 dBm & $\tau_{th}$ & 0.8 ms\\
		\hline
		$h^\textrm{u}$,$h^\textrm{d}$ & 10-tap Rayleigh fading channel & $\eta$ & 2 \\
		\hline
		$\sigma^2$ & 1 & $\beta$ & 2 kB \\
		\hline
	\end{tabular}\label{sys_param} \vspace{-0.55cm}
\end{table}
For simulations, $5$ agents are assigned to each environment and each agent utilizes the A3C algorithm to update its local policy network, and the target agent belongs to the Acrobot-v1 environment. 
We implement the A3C algorithm using Pytorch and the multiprocessing package. To avoid complexity and faster execution, we consider a fully connected DNN with two hidden layers and $\{128,128\}$ neurons for actor and critic networks. A \textit{tanh} activation function is used and an \textit{Adam} optimizer is considered for training the network. 
\begin{figure}[t!]
  \centering
    \includegraphics[width=7.3cm]{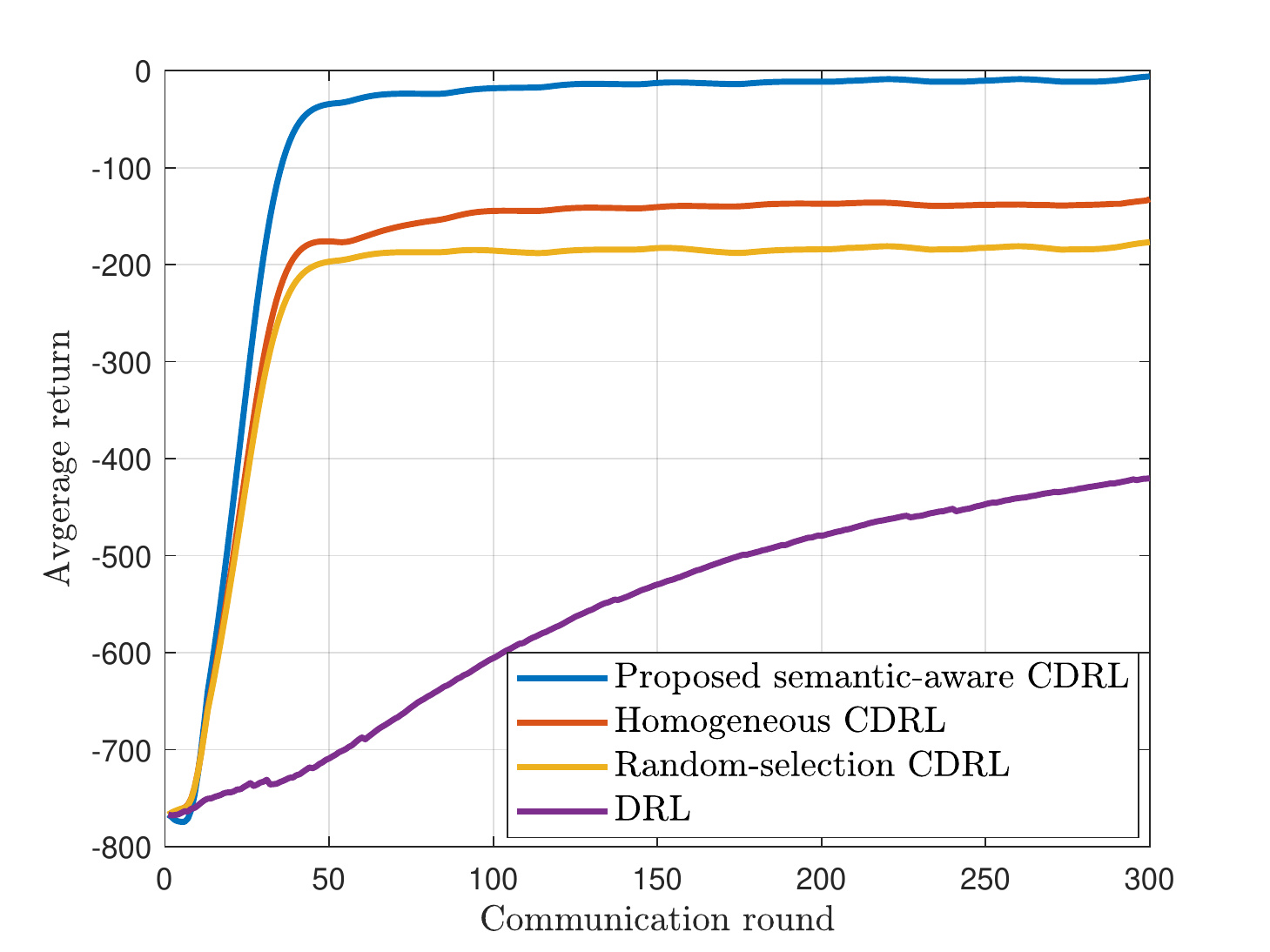}\vspace{-0.2cm}
    \caption{\small Performance comparison of CDRL algorithms.}\vspace{-0.2cm}
    \label{comp}
\end{figure}
The agents are randomly and uniformly distributed in the coverage area of the BS, modeled as a circular area with radius $250$ meters. To manage the data traffic of the proposed CDRL scheme, OFDMA transmissions are considered. Table \ref{sys_param} summarizes OFDMA parameters according to the 3GPP standards \cite{3gpp38}. 

Figure \ref{comp} compares the performance of the proposed semantic-aware HFDRL algorithm with three baseline methods: 1) Homogeneous CDRL, whereby the learning is performed only among the agents in the same environment, 2) Random-selection CDRL whereby the agent selection is random without accounting for similarity measures, and 3) conventional noncooperative DRL. The return values are averaged over sufficiently large number of runs. The results in Fig. \ref{comp} shows that the proposed method can yield up to $83\%$ higher final return compared to the baseline methods, thus, demonstrating the effectiveness of the proposed semantic-aware agent-selection scheme. Fig. \ref{comp} also shows that the proposed method can achieve a faster convergence compared to the baselines.   
That is due to the use of related agents with both structural and semantic relatedness, whereas homogeneous CDRL only relies on the structural similarity and ignores the semantics relatedness, and randomly selected CDRL ignores all similarity metrics.

\begin{figure}[t!]
  \centering
    \includegraphics[width=7.3cm]{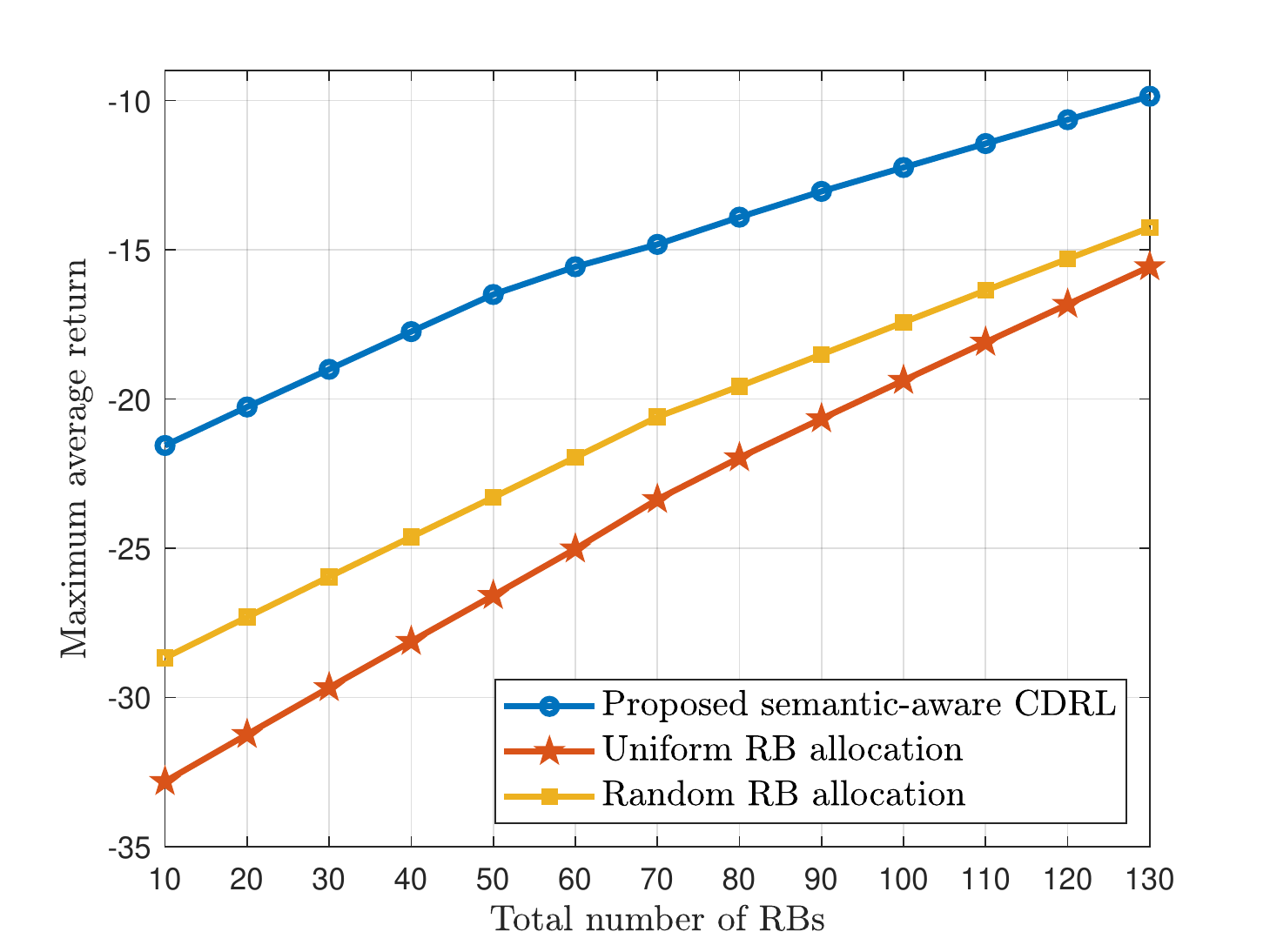}\vspace{-0.2cm}
    \caption{\small Impact of the total number of RBs on the system performance.}\vspace{-0.5cm}
    \label{NRB}
\end{figure}

Figure \ref{NRB} shows the maximum average return achieved after policy training versus the total number of RBs. 
Fig. \ref{NRB} shows that the proposed algorithm can yield up to $35\%$ and $27\%$ gains in terms of the average return, compared to the uniform and random RB allocation approaches, respectively. The results in Fig. \ref{NRB} also show that how the performance of the wireless network can affect the CDRL. In fact, as more wireless resources become available to agents, the transmission delays decrease which can lead to enhanced collaborative learning for the agents via the proposed HFDRL algorithm. In fact, the results in Fig. \ref{NRB} demonstrate the efficiency of the proposed algorithm for enabling CDRL over bandwidth-constrained wireless networks.  
\begin{figure}[t!]
  \centering
    \includegraphics[width=7.3cm]{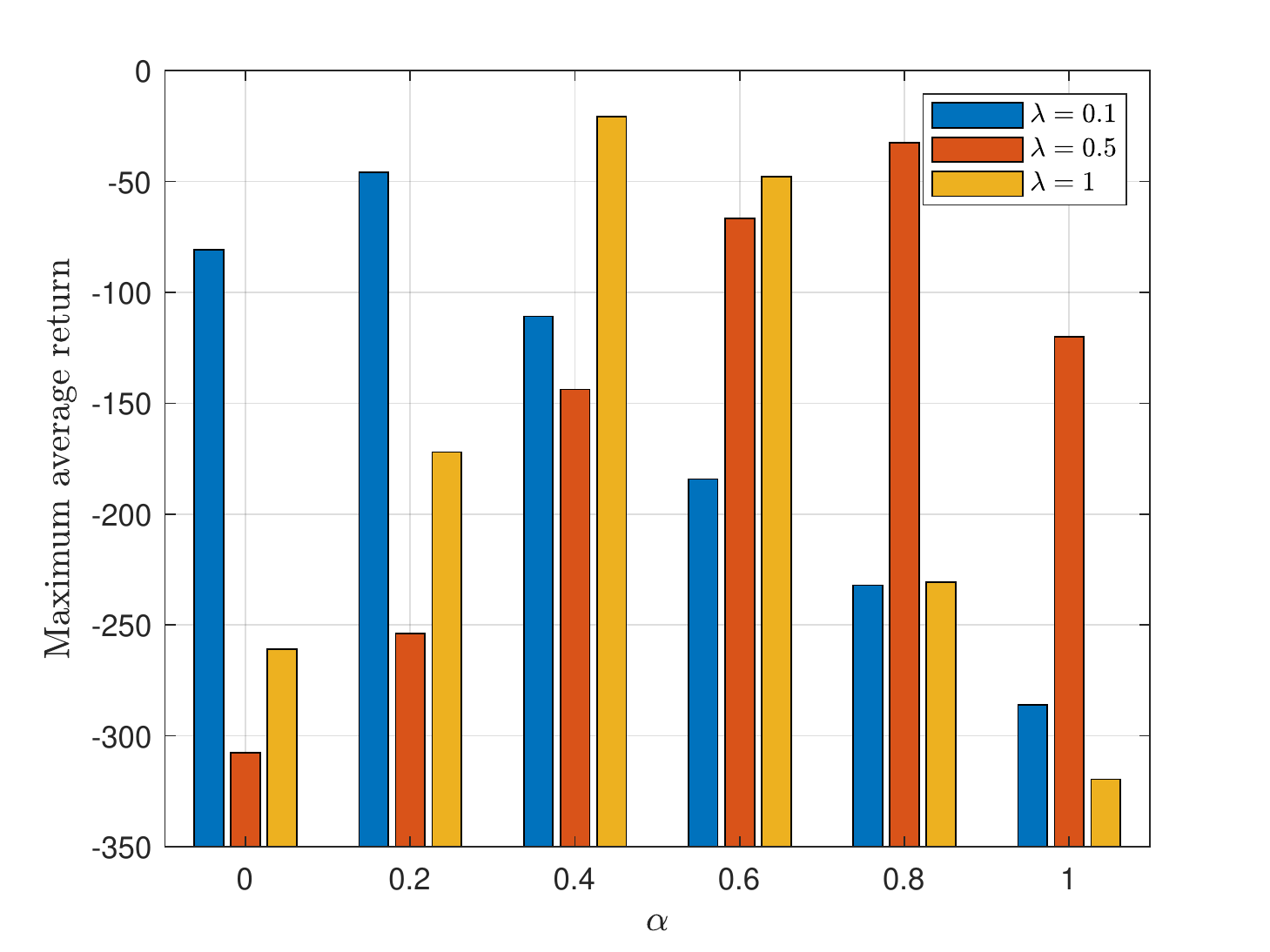}\vspace{-0.3cm}
    \caption{\small Impact of scalar variable $\alpha$ and $\lambda$ in the similarity measure on the system performance.}\vspace{-0.2cm}
    \label{alpha}
\end{figure}

Figure \ref{alpha} shows the maximum average return achieved after policy training versus the scalar $\alpha$ for various threshold values $\lambda$ used in the proposed agent selection scheme. The results in Fig. \ref{alpha} shed light on how to select $\lambda$ in proportion to $\alpha$ in order to achieve the highest average return. Here, note that selecting $\lambda=1$ implies a more strict criteria for allowing a source agent to participate in CDRL compared to smaller $\lambda$ values such as $\lambda=0.5$ and $\lambda=0.1$. 
Fig. \ref{alpha} demonstrates that using the proposed similarity measures in \eqref{mu} improves system performance by $83\%$ and $75\%$, respectively, as compared to using only semantic similarity ($\alpha=0$) or structural similarity ($\alpha=1$). \vspace{-0.2cm}
\section{Conclusions}\label{conclusion}
In this paper, we have developed a novel semantic-aware CDRL framework that enables efficient knowledge transfer among heterogeneous agents with semantically related tasks. To this end, we have formulated an optimization problem to find an optimal policy for a target agent, while accounting for both CDRL and wireless constraints. To solve this problem, we have developed a new HFDRL algorithm for selecting  the best subset of semantically-related DRL agents for collaboration. Accordingly, the training loss and wireless bandwidth allocation have been optimized jointly to train each agent within the time restriction of its real-time task. The simulation results have shown up to $83\%$ improvements in maximum rewards compared to the baseline methods. Further, the results have highlighted the importance of joint wireless resource optimization and policy training in CDRL, and have shown the efficiency of the proposed algorithm in presence of wireless bandwidth limitations. \vspace{0.0cm}

\def\baselinestretch{.92}
\bibliographystyle{IEEEbib}
\bibliography{Main}
\end{document}